\documentclass[aps,prl,twocolumn,superscriptaddress,groupedaddress]{revtex4}  % for review and submission
\usepackage{graphicx}  % needed for figures
\usepackage{dcolumn}   % needed for some tables
\usepackage{bm}        % for math
\usepackage{amssymb}   % for math
\usepackage[latin9]{inputenc}
\usepackage{amsmath}
\usepackage{subfigure}
\usepackage{amssymb}
\usepackage{graphicx}
\usepackage{amsfonts}%
\usepackage{dcolumn}% Align table columns on decimal point
\usepackage{bm}% bold math
\usepackage{hyperref}
\usepackage{nccmath}
\usepackage{lipsum}
\newenvironment{mpmatrix}{\begin{medsize}\begin{pmatrix}}%
		{\end{pmatrix}\end{medsize}}%

\begin{document}
	\title{\bf Systematic analysis for triple points in all magnetic symmorphic systems and symmetry-allowed coexistence of Dirac points and triple points}
%	\title{\bf Unexpected Weyl pairs emerge in Weyl-semimetal through continuous TRS phase changing among Dirac-semimetal Three-fold-degenerate-fermion-semimetal and Weyl-semimetal}
	\author{Chi-Ho Cheung}\email{f98245017@ntu.edu.tw}
	\affiliation{Graduate Institute of Applied Physics, National Taiwan University, Taipei 10617, Taiwan}
	\author{R. C. Xiao}
	\affiliation{Key Laboratory of Materials Physics, Institute of Solid State Physics, Chinese Academy of Sciences, Hefei 230031, China}
	\author{Ming-Chien Hsu}
	\affiliation{Department of Physics, National Sun Yat-Sen University}
	\author{Huei-Ru Fuh}
	\affiliation{Department of Chemical Engineering and Materials Science, Yuan Ze University, Taoyuan 33302, Taiwan}
	\author{Yeu-Chung Lin}
	\affiliation{Department of Physics, National Taiwan University, Taipei 10617, Taiwan}
	\author{Ching-Ray Chang}
	\affiliation{Department of Physics, National Taiwan University, Taipei 10617, Taiwan}

	\date{\today}

	\begin{abstract}
	%	In condensed matter physics, a recently discovered topological fermion-triple point has novel transport property and has no analogue of elementary fermionic particles in high energy physics. Hence, searching triple point materials for industrial application and extending the knowledge of this new-discovered topological fermion become important research items. 
		
		Similar to Weyl fermions, a recently discovered topological fermion ``triple point" can be generated from the splitting of Dirac fermion while the system has inversion symmetry (IS) breaking or time reversal symmetry (TRS) breaking. Inducing triple points in IS breaking symmorphic systems have been well studied, whereas in TRS breaking symmorphic systems have not yet. In this work, we extend the theory of searching triple points to all symmorphic magnetic systems. We list all $k$ paths of all symmorphic systems which allow the existence of triple points. With this systematic study, we also find out that the coexistence of Dirac points and triple points is symmetrically allowed in some particular symmetric systems. Our works will not only be helpful for searching triple points but also extend the knowledge of such a new topological fermion.
	\end{abstract}
	
	\maketitle
	
	\section{I. Introduction}
		Over the past few decades, topology has been emerging in condensed matter physics. The development started from quantum Hall effect\cite{in1-1, in1-2} which is the quantum-mechanical version of Hall effect. The second stage of development is quantum anomalous Hall effect\cite{in2-1, in2-2, in2-3, in2-4} which is a quantum Hall effect without external magnetic field. The third stage of development is quantum spin Hall effect\cite{in3-1, in3-2, in3-3, in3-4, in3-5, in3-6} which is a quantum Hall effect without breaking time reversal symmetry (TRS). Analogous to quantum spin Hall effect which pumps spin, there are topological crystalline insulators\cite{in4-1, in4-2, in4-3, in4-4} which can pump the eigenvalues of mirror symmetry. All these four topological phenomena are insulating in bulk band, but have topologically protected surface state which is conducting.

Besides looking for topological phenomena in bulk insulating materials, scientists also look for topological phenomenon in bulk metallic materials. Recently, topological metals such as Dirac semimetal\cite{in5-1, in5-2, in5-3, in5-4},  Weyl semimetal\cite{in6-1, in6-4, in6-3, in6-2} and triple point semimetal\cite{in7-1, in7-2, in7-3, in7-4, in7-5, in7-6, in7-7, in7-8, in7-9, in7-10, in7-11, in7-12} have been discovered. These topological metals have topologically protected surface state just like those quantum Hall effects. No matter metallic in bulk or insulating in bulk, as long as their surface state are topologically protected, they can be promising candidates for electronic devices or even spintronic devices. Thus they can be valuable for industrial applications. On the other hand, topological metal provides a different playground and relatively lower price to search for those elementary particles described by relativistic quantum field theory. Since topological metal is valuable for both academic research and industrial applications, it has drawn a lot of attention in recent years.

One of the topological metals hosting a quasiparticle analogue of an elementary particle is Dirac semimetal. The earliest found Dirac semimetal is $Na_{3}Bi$\cite{in5-3}. $Na_{3}Bi$ has both IS and TRS, thus all bands at every $k$ points in the Brillouin zone are at least doubly degenerate. While any doubly degenerate band linearly crossing over another doubly degenerate band at a $k$ point, a four-fold degenerate Dirac point is formed. Such a Dirac point can be an analogue of the Dirac fermion which described by relativistic quantum field theory in high energy physics.

In high energy physics, breaking TRS or IS can cause Dirac fermion splitting into Weyl fermions. In condensed matter physics, bands can be non-degenerate while system does not have TRS and IS. When a non-degenerate band linearly crossing over another non-degenerate band at a $k$ point, a two-fold degenerate Weyl point is formed. Such a Weyl point can be an analogue of Weyl fermion in high energy physics too.

However, in condensed matter physics, fermions in crystal are constrained by magnetic space group symmetries rather than by Lorentz invariance. This gives rise to the uncertainty that doubly degenerate bands may or may not split while TRS or IS is broken. In this paper, we will discuss a new fermion-triple point which has no counterparts in high-energy physics and can be formed by a non-degenerate band linearly crossing over another doubly degenerate band at a $k$ point. In general, the formations of triple points can be caused by nonsymmorphic or symmorphic magnetic space groups symmetries, but as we emphasize in the title, we only discuss those triple points which caused by symmorphic magnetic space group symmetries.

If Dirac fermions in condensed matter must has TRS and IS just like the Dirac fermions in high energy physics, then it cannot coexist with triple point which cannot exist in a system with both TRS and IS. However, recent research shows that Dirac fermions in condensed matter can exist in a system without TRS $\cdot$ IS\cite{in8-1, in8-2}. This gives rise to the possibility of finding several systems which have two k paths with two different symmetry groups, one allows the existence of Dirac points while another one allows the existence of triple points.

%Furthermore, the degeneracies of bands depend on the symmetry group of the k point, and different k points in a same system may have different symmetry groups. This gives rise to the possibility of finding a several systems which have two k paths with two different symmetry groups, one allows the existence of Dirac points while another one allows the existence of triple points.

We organize this paper as follows. In section II, we review the condition of forming triple points by discussing a special case. In section III, we generalize this condition to all magnetic point groups and list all possible k paths of all possible symmorphic systems which allow the existence of triple points. In section IV, we point out that the coexistence of Dirac points and triple points is symmetrically allowed in some particular symmetric systems. In section V, we summarize the contributions of this paper.

\section{II. The condition of forming triple points}
Similar to Weyl fermions in high energy physics, triple points in condensed matter physics can be split from Dirac fermions while TRS or IS of the system is broken. It is well known that Dirac fermions can exist in a system which has both TRS and $D_{6h}$ point group symmetry. In this section, we are going to use this system as an example to show how triple points split from Dirac fermions and point out the necessary condition of forming triple points.

$D_{6h}$ point group includes $C_{3z}$, $C_{2z}$, $M_{x}$ and IS. Since the system has TRS and IS, all bands have spin degeneracy at any $k$ point. As Dirac fermions are a crossing point of two 2-fold degenerate bands, Dirac fermion is a point of 4-fold degeneracy.

If all bands have spin degeneracy at any $k$ point, triple point cannot be formed (triple point is a point of 3-fold degeneracy). Thus TRS or IS must be broken to induce triple points. However, at a high symmetry $k$ point/path/plane, TRS and IS are not the only symmetries which protect the degeneracy of bands. Therefore, considering other crystal symmetries is needed.

To be more specific, we assume the irreducible representations of the bands which form the Dirac points are $\overline{E}_{1g/u}$ and $\overline{E}_{3g/u}$. With the irreducible representations, the matrix forms of the symmetry operators are as follows:
	\begin{equation} \label{eq:1}
		%\begin{array}
		\begin{split}
			\begin{array}{lcl}
				for \ basis:
				 \overline{E}_{1g/u} \\
				%			\begin{bmatrix}
				%			S^{\pm}_{\frac{1}{2}},+\frac{1}{2} \\
				%			S^{\pm}_{\frac{1}{2}},-\frac{1}{2} 
				%			\end{bmatrix} \\
				\ \ \ \ \ TRO \ \ \ \ \ \ \ \ \ \ \  IS  \ \ \ \ \ \ \ \ \  M_{x}  \ \ \ \ \ \  C_{2z} \ \ \ \ \ \ \ \ \  C_{3z}  \\ 
				{\begin{matrix}
						\pm\begin{bmatrix}
			0 & -1 \\ 
			1 & 0
						\end{bmatrix}K &  \pm\begin{bmatrix}
				1 & 0 \\ 
				0 & 1
					\end{bmatrix} &  \pm\begin{bmatrix}
					0 & i \\ 
					i & 0
				\end{bmatrix} &  \begin{bmatrix}
						i & 0 \\ 
						0 & -i
			\end{bmatrix} &  \begin{bmatrix}
					e^{\frac{i\pi}{3}} & 0 \\ 
					0 & e^{\frac{-i\pi}{3}} 
		\end{bmatrix}
	\end{matrix}}, \\
	\\
	for \ basis:
     \overline{E}_{3g/u} \\
	%    \begin{bmatrix}
	%    P^{\pm}_{\frac{3}{2}},+\frac{3}{2} \\
	%    P^{\pm}_{\frac{3}{2}},-\frac{3}{2}
	%    \end{bmatrix} \\
\ \ \ \ \ TRO \ \ \ \ \ \ \ \ \ \ \ \ \ \  IS  \ \ \ \ \ \ \ \ \ \ \  M_{x}  \ \ \ \ \ \  C_{2z} \ \ \ \ \ \ \ \  C_{3z}  \\ 
	{\begin{matrix}
		\pm\begin{bmatrix}
0 & 1 \\ 
-1 & 0
		\end{bmatrix}K &  \mp\begin{bmatrix}
	-1 & 0 \\ 
	0 & -1
		\end{bmatrix} &  \mp\begin{bmatrix}
		0 & i \\ 
		i & 0
		\end{bmatrix} &  \begin{bmatrix}
		-i & 0 \\ 
		0 & i
		\end{bmatrix} &  \begin{bmatrix}
		e^{i\pi} & 0 \\ 
		0 & e^{-i\pi} 
		\end{bmatrix}
		\end{matrix}},
\end{array}
\end{split}
%\end{array}
\end{equation}
%\end{multline}
%\end{widetext}
%\end{center}
where $TRO$ is the operator of TRS and $K$ is complex conjugate operator.

%With $D_{6h}$ point group symmetry and TRS, Dirac points locate at $\Gamma-Z$ axis which is the principle axis of the crystal.

If a symmetry operator S (S can be a unitary or an anti-unitary operator) acts on a $k_{h}$ vector, such that $Sk_{h}=k_{h}+nG$, where G is any reciprocal lattice vector and n is any integer number, then all such symmetry operators form the little group of $k_{h}$.

Firstly, we only consider the unitary subgroup of the little group of $k_{h}$. Hamiltonian $H(k_{h})$ has to commute with all the symmetry operators of the unitary subgroup of the little group of $k_{h}$. If any symmetry operators of this unitary subgroup does not commute with each other in a subspace of the Hilbert space, then $H(k_{h})$ has to be degenerate in this subspace, otherwise $H(k_{h})$ cannot commute with all the symmetry operators of the unitary subgroup simultaneously.

Furthermore, those anti-unitary symmetry operators of the little group of $k_{h}$ could cause extra degeneracy. In symmorphic system, in order to consider all the symmetry operators of the little group of $k_{h}$, we have to treat the little group as a magnetic point group rather than to treat it as the original point group, no matter the system does or does not have any magnetic moment. If the system does not have any magnetic moment, then it has TRS. Thus the symmetry group of the system is one of the grey groups of the 122 magnetic point groups. The little group of $k_{h}$ of this system is a subgroup of the grey group. Therefore, the little group of $k_{h}$ of a paramagnetic system could be any magnetic point group. All 122 magnetic point groups can be classified into three types: 32 ordinary point groups, 32 ``grey" point groups and 58 ``black and white" magnetic point groups. The degeneracies of the ordinary point groups have been discussed hereinabove. The extra degeneracies that caused by the TRS of any grey point groups are known to be Kramers degeneracy which has been well discussed too\cite{con1}. The extra degeneracies which caused by those anti-unitary symmetry operators of any black and white magnetic point groups are discussed in the Appendix of this paper.

In the system with $D_{6h}$ and TRS, any $k$ point on $\Gamma-Z$ axis-$k_{z}$ is invariant under $C_{3z}$, $C_{2z}$ rotation or $M_{x}$ reflection or (TRO$\cdot$IS) operation, thus the symmetry group of $\Gamma-Z$ axis is $6/m'mm$ which is a black and white magnetic point group. According to Eq. \ref{eq:1}, $C_{2z}$ does not commute with $M_{x}$ in both $\overline{E}_{1g/u}$ and $\overline{E}_{3g/u}$. Furthermore, according to Table. \ref{tab:S1}, the anti-unitary operators in $6/m'mm$ do not cause any extra degeneracy. Thus, $\overline{E}_{1g/u}$ and $\overline{E}_{3g/u}$ are both 2-fold degeneracy along the $k_{z}$ path. Besides, $\overline{E}_{1g/u}$ and $\overline{E}_{3g/u}$ are two different irreducible representations in $k_{z}$ path, so any coupling between these two representations (bands) are forbidden. Hence, there will be no gap opening when these two bands come across each other at $k_{z}$ path. Therefore, under such symmetry condition, a linear crossing between two 2-fold degenerate bands is allowed and so is the 4-fold degenerate Dirac point.

The symmetry condition of allowing the existence of Dirac points can be streamlined and generalized as follows: Dirac points can exist at a $k$ path whose symmetry group has two or more than two 2-dimensional double group irreducible representations.  

\begin{figure}
	\centering
	\includegraphics[width=8.5cm]{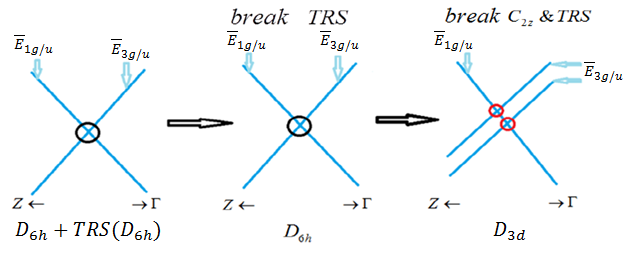}
	\caption{The schematic figure of the symmetry reduction processes of the system. After the system symmetry has been reduced from $D_{6h}+TRS(D_{6h})$ to $D_{6h}$, both $\overline{E}_{1g/u}$ and $\overline{E}_{3g/u}$ do not split. When the symmetry of the system becomes $D_{3d}$, $\overline{E}_{1g/u}$ remains to be a 2-fold degeneracy whereas $\overline{E}_{3g/u}$ splits into two non-degenerate bands. The Dirac points are marked in black circles and the triple points are marked in red circles.
	}\label{Fig: 3}
\end{figure}

If the TRS of the system is broken, the symmetry group of $k_{z}$ path is reduced from $6/m'mm$ to $C_{6v}$. Since the 2-fold degeneracy of $\overline{E}_{1g/u}$ and of $\overline{E}_{3g/u}$ remain being protected by $C_{2z}$ and $M_{x}$, the Dirac points on $k_{z}$ path do not split just because of TRS breaking. If $M_{x}$ symmetry is chosen for further symmetry breaking, all symmetry operators of the little group of $k_{z}$ path commute with each other. Both $\overline{E}_{1g/u}$ and $\overline{E}_{3g/u}$ will be split. If we want to induce triple points, breaking $M_{x}$ symmetry is not an option. If $C_{2z}$ is chosen for the further symmetry breaking, the symmetry group of the little group of $k_{z}$ path becomes $C_{3v}$. All symmetry operators in $\overline{E}_{3g/u}$ commute with each other, the symmetry operators in $\overline{E}_{1g/u}$ do not commute with each other. Thus $\overline{E}_{1g/u}$ remains to be a 2-fold degeneracy whereas $\overline{E}_{3g/u}$ splits into two non-degenerate bands. On $k_{z}$ path, since $C_{3z}$ symmetry can prevent any coupling between $\overline{E}_{1g/u}$ and $\overline{E}_{3g/u}$, these representations still belong to different irreducible representations. Therefore, the crossing point will not be gapped. Thus each Dirac point will split into two triple points when the $C_{2z}$ and TRS are broken. The variations of band structures and of system symmetry are shown in Fig. \ref{Fig: 3}.

This physical phenomenon will be further clarified if we use the $k \cdot P $ expansion and method of invariants to calculate the Hamiltonian around $\Gamma$ point for $k_{z}$ path:
		\begin{equation} \label{eq:2}
			\begin{array}{lcl}
				H_{D_{3d}}(k_{z})=\varepsilon_{0}(k_{z})+ \\
				\begin{mpmatrix}
					C_{0}-C_{1}k^{2}_{z} & 0 & 0 & 0 \\
					0 & C_{0}-C_{1}k^{2}_{z} & 0 & 0 \\
					0 & 0 & -C_{0}+C_{1}k^{2}_{z} & D \\
					0 & 0 & D & -C_{0}+C_{1}k^{2}_{z} \\
				\end{mpmatrix},
			\end{array}
		\end{equation}
this is the Hamiltonian for the system with $D_{3d}$ symmetry and without TRS; the expansion is only up to the first order of $k$ for off-diagonal matrix elements and up to the second order of $k$ for diagonal matrix elements; $\varepsilon_{0}(k_{z})=A_{0}+A_{1}k_{z}^{2}$. $C_{0}$ and $C_{1}$ are real positive $k$ independent coefficients. $A_{0}$, $A_{1}$ and $D$ are real $k$ independent coefficients.

When the $C_{3z}$ operator acts on the effective Hamiltonian $C_{3z}H_{D_{3d}}(k)C^{-1}_{3z}$, according to the $C_{3z}$ symmetry operator of Eq. \ref{eq:1}, a phase factor $e^{\frac{2i\pi}{3}}$ or $-e^{\frac{i\pi}{3}}$ will be generated on the matrix elements $H_{12}$, $H_{21}$ and the matrix elements of off-diagonal block. Therefore these matrix elements must have $k_{+}$ or $k_{-}$ ($k_{\pm}=k_{x}\pm ik_{y}$) factor to match the $C_{3z}$ symmetry condition $C_{3z}H_{D_{3d}}(k)C^{-1}_{3z}=H_{D_{3d}}(C^{-1}_{3z}k)$. If only considering the Hamiltonian of $\Gamma-Z$ axis, all these matrix elements become zero. This explains hereinabove mentioned: on $k_{z}$ path, $C_{3z}$ symmetry can prevent any coupling between $\overline{E}_{1g/u}$ and $\overline{E}_{3g/u}$, such that these representations belong to different irreducible representations. Furthermore breaking $C_{2z}$ symmetry and TRS induces $D(\tau_{0}\sigma_{x}-\tau_{z}\sigma_{x})/2$ ($\tau$ is the Hilbert space of combining $\overline{E}_{1g/u}$ and $\overline{E}_{3g/u}$; $\sigma$ is the Hilbert space within $\overline{E}_{1g/u}$ or within $\overline{E}_{3g/u}$). This term splits $\overline{E}_{3g/u}$ to two 1-dimensional irreducible representations. Thus the 4-fold degenerate Dirac point splits into two 3-fold degenerate triple points.

Base on the above analysis, the symmetry condition of allowing the existence of triple points can be streamlined as follows: triple points only can exist at a k path whose symmetry group contains both 1-dimensional and 2-dimensional double group irreducible representations.

\section{III. Triple points in all symmorphic systems}
In this section, we are going to find out all possible $k$ paths of all possible symmorphic systems which allow the existence of triple points.

In symmorphic system, symmetry group of any k points is one of the magnetic point groups. So the first step is to check those 122 types magnetic point group and list all magnetic point groups which can match the symmetry condition of allowing the existence of triple points.

\begin{figure}[b]
		\centering
			\includegraphics[width=5.5cm]{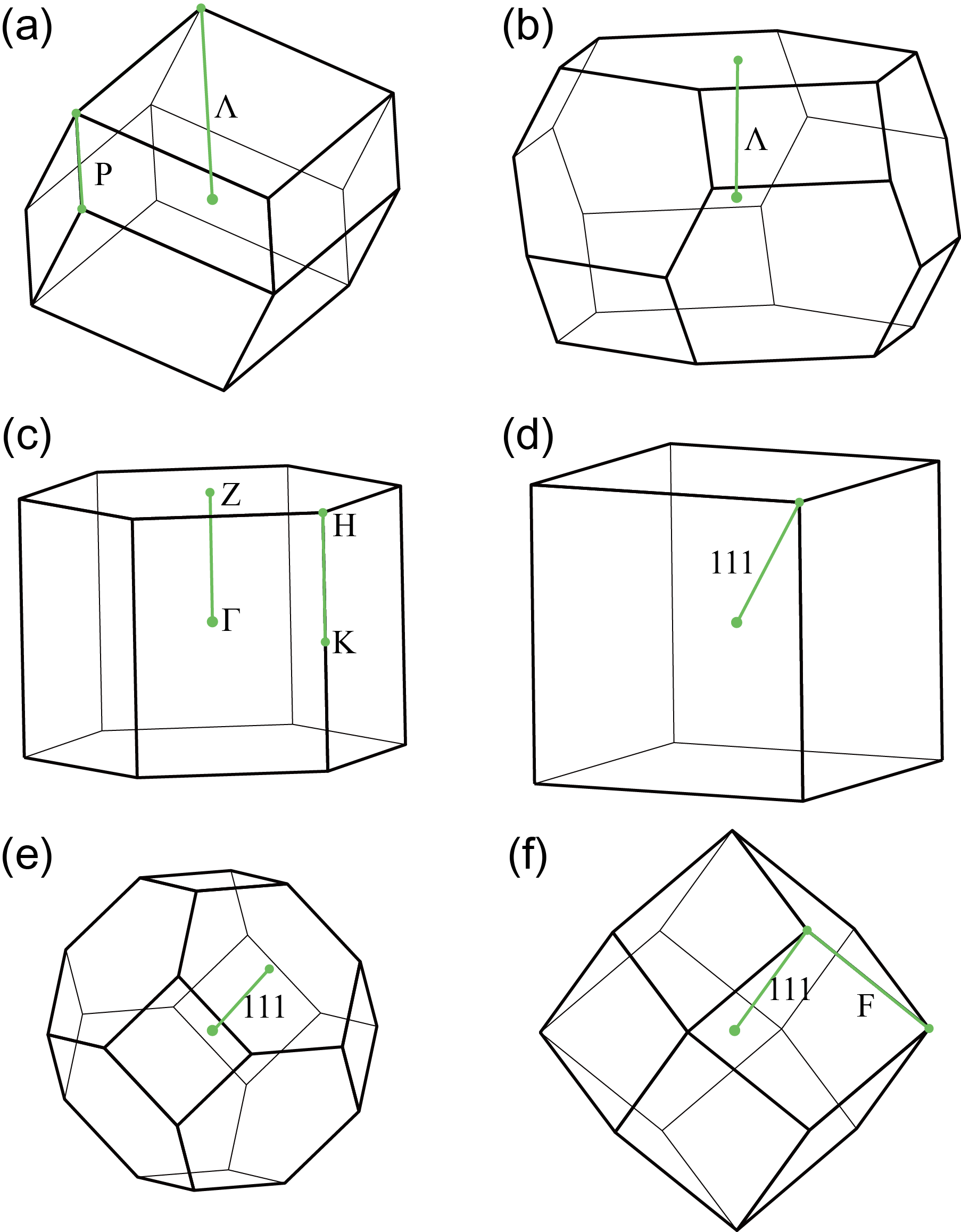}
		\caption{The Brillouin zone for (a) trigonal with $a > \sqrt{2} c$, (b) trigonal with $a < \sqrt{2} c$, (c) hexagonal, (d) cubic $P$, (e) cubic $F$ and (f) cubic $I$. All the $k$ paths, which contain $C_{3}$ symmetry, are marked in green. Other Brillouin zones, which cannot contain $C_{3}$ symmetry, are omitted.
		}\label{Fig: 1}
\end{figure}

Among 32 types of ordinary point groups, there are 3 types, namely $C_{3v}$, $D_{3}$ and $D_{3d}$, of point groups containing both 1-dimensional and 2-dimensional double group irreducible representations. In all the Brillouin zone of the 14 types Bravais lattice, only 6 types of $k$ path contain $C_{3}$ symmetry (there is no $k$ plane contains $C_{3}$ symmetry). These 6 types of $k$ path are $\Lambda$ and $P$ of trigonal; $\Gamma-Z$ and $K-H$ of hexagonal; $1 1 1$ direction of cubic $P$, cubic $F$ and cubic $I$; F of cubic $I$ (since there is no unified $k$ path symbol, the Brillouin zones are demonstrated in Fig. \ref{Fig: 1} to define the 6 symbols which are used for marking the 6 types of $k$ path). These 6 types of $k$ path only can contain $C_{3v}$, none of them can contain $D_{3}$ or $D_{3d}$. Therefore, among the ordinary point groups only $C_{3v}$ can match the symmetry condition of allowing the existence of triple points.

Since operating TRO on $k$ is to change the sign of $k$, only $k$ points, not $k$ paths, allow grey point group to be their symmetry group. Thus, if the symmetry group of $k$ is a grey point group, triple point cannot exist on this $k$.

All black and white magnetic point groups contain a set of unitary operators which form a unitary subgroup (one of the ordinary point group), and this unitary subgroup has a set of double group irreducible representations. The rest of the operators of the black and white magnetic point group are anti-unitary operators. These anti-unitary operators cannot generate any double group irreducible representation, but they can further degenerate the ordinary double group irreducible representations (as shown in Appendix). The further degeneracies could allow the existence of triple points, whereas the ordinary double group irreducible representations forbid that. Or contrarily, the further degeneracies forbid the existence of triple points, whereas the ordinary double group irreducible representations allow that.

Among all the 58 types of black and white magnetic point groups, 17 types have further degeneracies (as shown in Table. \ref{tab:S1}). Among the 17 types, only $-6'$ allows the existence of triple points, whereas the double group irreducible representations of its unitary subgroup forbid that. All the other 16 types forbid the existence of triple points due to: having the element-TRS $\cdot$ IS or having no 1-dimensional double group irreducible representation or no $k$ path belonging to the black and white magnetic point group. Among the 16 types, $-3'm$ is the one that forbids the existence of triple points, whereas the double group irreducible representations of its unitary subgroup (its unitary subgroup is $C_{3v}$) allow that. For later discussion, it would be important to notice that $-3'm$ has the element of $-1'$.

Next step is to look for $k$ paths, in all symmorphic systems, which allow existence of triple points. We can directly look for these $k$ paths in all the symmorphic magnetic space groups. However, there are too many symmorphic magnetic space groups. Thus we choose to analyze magnetic point group of the symmorphic magnetic space group first. In this way a large number of unqualified magnetic space groups can be excluded. Then we contrast those qualified magnetic point group systems with their magnetic space group symbols.

As mentioned above, there are two kinds of $k$ paths allowing the existence of triple points: the first kind is the $k$ paths which belong to $-6'$ black and white magnetic point group; the second kind is the $k$ paths whose unitary subgroup belongs to $C_{3v}$ while the $k$ paths do not have $-1'$ symmetry. We will search these two kinds of $k$ paths separately.

Firstly, we search the $k$ paths which belong to $-6'$. If the Brillouin zone of a system contains the $-6'$ $k$ path, the crystal symmetry (directions of magnetic moments are not counted for crystal symmetry) of the system must contains $C_{3h}$ symmetry. According to the subgroup decomposition of the 32 point groups in Ref. \cite{con1}, the crystal symmetry of the system contains $C_{3h}$ symmetry only if the crystal symmetry of the system is $D_{6h}$ or $D_{3h}$ or $C_{6h}$ or $C_{3h}$. The Bravais lattice of these four kinds of crystal symmetry is Hexagonal. In the Brillouin zone of Hexagonal Bravais lattice, only $\Gamma-Z$ can contain all the symmetry elements of $-6'$. Thus we only need to determine whether the $\Gamma-Z$ of all magnetic point group systems of $D_{6h}$, $D_{3h}$, $C_{6h}$ and $C_{3h}$ belong to $-6'$. The results are shown at Table. \ref{tab:1}.

\begin{table}[h]
	\caption{The list for searching first kind of $k$ paths. The first column is the label of magnetic point group systems. Answer of ``Does $\Gamma-Z$ of the system belong to $-6'$?" is given in second column.}
	\begin{tabular}{c | c}
		\hline
		\hline
		Label of magnetic point group systems & Does $\Gamma-Z$ belong to \\ & $-6'$? \\
		\hline
		($C_{6h}$) 6/m &  no         \\
		6/m1' &  no       \\
		6'/m' &  yes         \\
		6'/m &  no         \\
		6/m' &  no         \\
		($D_{3h}$) -6m2 &  no       \\
		-6m21' &  no       \\
		-6m'2' &  no         \\
		-6'm2' &    no     \\
		-6'm'2&       yes          \\
		 ($D_{6h}$) 6/mmm & no       \\
		  6/mmm1'  &    no              \\
		 6/m'mm &    no    \\
		 6/mm'm' &   no          \\
		 6/m'm'm' & no  \\
		 6'/mmm' &  no  \\
		 6'/m'mm' & no  \\
		 ($C_{3h}$) -6 &  no \\
		 -61'  &  yes  \\
		 -6'   &  yes  \\
		\hline
		\hline
	\end{tabular}
	\label{tab:1}
\end{table}

Secondly, we search the $k$ paths whose unitary subgroup belongs to $C_{3v}$ while the $k$ paths do not have $-1'$ symmetry. If the Brillouin zone of a system contains such $k$ path, the symmetry group of the system must contain $C_{3v}$ symmetry. According to the subgroup decomposition of the 32 point groups in Ref. \cite{con1}, the symmetry group of the system contains $C_{3v}$ symmetry only if the crystal symmetry of the system is one of the seven symmetries, namely $C_{3v}$, $T_{d}$, $O_{h}$, $C_{6v}$, $D_{3d}$, $D_{3h}$ and $D_{6h}$. Furthermore, as (TRO $\cdot$ IS) acting on any $k$ is equal to $k$, any $k$ point in the Brillouin zone contains (TRO $\cdot$ IS) symmetry if and only if the system contains (TRO $\cdot$ IS) symmetry. Thus we can use two symmetry conditions to filter most of the magnetic point groups which belong to the seven crystal symmetry: 1. the system must contains $C_{3v}$ symmetry; 2. the system must not contains (TRO $\cdot$ IS) symmetry. Besides, the $k$ paths which contain $C_{3v}$ symmetry must contain $C_{3}$ symmetry. Hence, as mentioned above, the $k$ path whose unitary subgroup is $C_{3v}$ symmetry must be one of the following: $\Lambda$ and $P$ of trigonal; $\Gamma-Z$ and $K-H$ of hexagonal; $1 1 1$ direction of cubic $P$, cubic $F$ and cubic $I$; F of cubic $I$. All we need to do is to determine whether the unitary subgroup, of these 6 types of $k$ paths in the Brillouin zone of filtered magnetic point group systems, is $C_{3v}$ symmetry. The determining processes and results are given in Table. \ref{tab:2}. Notice that for the Brillouin zone of trigonal Bravais lattice, the $k$ path-$P$ appears only if lattice constants fulfill the condition of $a > \sqrt{2} c$. Sometimes, the $k$ path-$K-H$ of Hexagonal Bravais lattice is not contained in the mirror plane of $C_{3v}$ symmetry, such that $K-H$ does not contain $C_{3v}$ symmetry. Hence, we need the magnetic space group to determine whether the existence of triple points is allowed on this $k$ path. Thus, some answers in the fifth column of Table. \ref{tab:2} are ``MSG is needed". Combining Table. \ref{tab:1} and Table. \ref{tab:2}, then contrasting with magnetic space group, we can get all $k$ paths of all symmorphic systems which allow the existence of triple points (as shown in Table. \ref{tab:3}).

\begin{widetext}
\begin{center}
\begin{table}[h]
	\caption{The list for searching second kind of $k$ paths. The first column is the label of magnetic point group systems. The second column shows whether the system be filtered out by the symmetry condition of containing $C_{3v}$ while does not contain $-1'$. The third column is the class of Brillouin zone. The fourth column is the label of $k$ path. The fifth column shows that whether the $k$ path allows the existence of triple points. Sometimes, the $k$ path-$K-H$ of Hexagonal Bravais lattice is not contained in the mirror plane of $C_{3v}$ symmetry and consequently $K-H$ does not contain $C_{3v}$ symmetry. Hence, we need the magnetic space group to determine whether the existence of triple points is allowed on this $k$ path. Thus, some answers in the fifth column are ``MSG is needed". For the Brillouin zone of trigonal Bravais lattice, the $k$ path-$P$ appears only if lattice constants fulfill the condition of $a > \sqrt{2} c$.}
	\begin{tabular}{c | c | c | c | c}
		\hline
		\hline
		symmetry of the system & contains $C_{3v}$ or not? contains $-1'$ or not? & Brillouin zone & $k$ path & the existence of triple points \\
		\hline
		($C_{3v}$) 3m & contains $C_{3v}$, but no $-1'$ & trigonal & $\Lambda$ & yes \\
		 & & & P & yes \\
		 & & Hexagonal & $\Gamma-Z$ & yes \\
		 & & & $K-H$ & MSG is needed \\
		 3m1' & contains $C_{3v}$, but no $-1'$ & trigonal & $\Lambda$ & yes \\
		& & & P & yes \\
		& & Hexagonal & $\Gamma-Z$ & yes \\
		& & & $K-H$ & MSG is needed \\		 
		3m' & no $C_{3v}$ & & & \\
		($T_{d}$) -43m  & contains $C_{3v}$, but no $-1'$ & cubic P & 1 1 1 & yes \\
		 & & cubic F & 1 1 1 & yes \\
		 & & cubic I & 1 1 1 & yes \\
		 & & & F & yes \\
		 -43m1'  & contains $C_{3v}$, but no $-1'$ & cubic P & 1 1 1 & yes \\
		 & & cubic F & 1 1 1 & yes \\
		 & & cubic I & 1 1 1 & yes \\
		 & & & F & yes \\
		-4'3m' & no $C_{3v}$ & & & \\
		($O_{h}$) m-3m & contains $C_{3v}$, but no $-1'$ & cubic P & 1 1 1 & yes \\
		 & & cubic F & 1 1 1 & yes \\
		 & & cubic I & 1 1 1 & yes \\
		 & & & F & yes \\
		m-3m1' & contains $-1'$ & & &   \\
		m'-3'm & contains $-1'$ & & &   \\
		m-3m'  & no $C_{3v}$ & & &   \\
		m'-3'm' & contains $-1'$ & & &   \\
		($C_{6v}$) 6mm & contains $C_{3v}$, but no $-1'$   & Hexagonal & $\Gamma-Z$ &   no    \\
		&  &  & $K-H$ & yes \\
		6mm1' & contains $C_{3v}$, but no $-1'$   & Hexagonal & $\Gamma-Z$ &   no    \\
		&  &  & $K-H$ & yes \\
		6'mm' &  contains $C_{3v}$, but no $-1'$   & Hexagonal & $\Gamma-Z$ &   yes    \\
		&  &  & $K-H$ & MSG is needed \\
		6m'm' & no $C_{3v}$ & & & \\
		
		($D_{3d}$) -3m & contains $C_{3v}$, but no $-1'$ & trigonal & $\Lambda$ & yes \\
		& & & P & yes \\
		& & Hexagonal & $\Gamma-Z$ & yes \\
		& & & $K-H$ & MSG is needed \\
		-3m1' & contains $-1'$        & & &        \\
		-3'm & contains $-1'$        & & &        \\
		-3'm' & no $C_{3v}$  & & &        \\
		-3m' & no $C_{3v}$  & & &        \\
		
		($D_{3h}$) -6m2 &  contains $C_{3v}$, but no $-1'$   & Hexagonal & $\Gamma-Z$ & yes    \\
		       &  &  & $K-H$ & MSG is needed \\
		-6m21' & contains $C_{3v}$, but no $-1'$   & Hexagonal & $\Gamma-Z$ &   yes    \\
		       &  &  & $K-H$ & MSG is needed \\
		-6m'2' &   no $C_{3v}$  & & &        \\
		-6'm2' &   contains $C_{3v}$, but no $-1'$    &  Hexagonal  & $\Gamma-Z$ &  yes  \\
		       &  &  & $K-H$ & MSG is needed \\
		-6'm'2 &     no $C_{3v}$      & & &       \\
		($D_{6h}$) 6/mmm & contains $C_{3v}$, but no $-1'$  &  Hexagonal & $\Gamma-Z$ &   no    \\
		       &  &  & $K-H$ & yes \\
		6/mmm1'  &    contains $-1'$        & & &        \\
		6/m'mm &    contains $-1'$   & & &    \\
		6/mm'm' &   no $C_{3v}$     & & &       \\
		6/m'm'm' & contains $-1'$ & & &   \\
		6'/mmm' & contains $-1'$  & & &   \\
		6'/m'mm' & contains $C_{3v}$, but no $-1'$ &  Hexagonal & $\Gamma-Z$ & yes  \\
		       &  &  & $K-H$ & MSG is needed \\
		\hline
		\hline
	\end{tabular}
	\label{tab:2}
\end{table}
\end{center}
\end{widetext}

\begin{table}[h]
	\caption{The list of all $k$ paths of all symmorphic systems which allow the existence of triple points. The first column is the label of magnetic point group systems which allow the existence of triple points. The second column are the $k$ paths of the system which allow the existence of triple points. The third column is the kind of the $k$ path. The first kind of $k$ paths is from Table. \ref{tab:1} and the second kind is from Table. \ref{tab:2}.}
	\begin{tabular}{c | c | c}
		\hline
		\hline
		Label of magnetic  & $k$ path & the kind of $k$ path \\
		point group systems &  &  \\
		\hline
		P3m1 &  $\Gamma-Z$   &   first kind   \\
		P3m11' & $\Gamma-Z$ & first kind \\
		P31m &  $\Gamma-Z$    &  first kind \\
		 & $K-H$ & first kind \\
		P31m1' & $\Gamma-Z$    &  first kind \\
		& $K-H$ & first kind \\
		R3m &   $\Lambda$   &   first kind  \\
		  & P & first kind \\
		R3m1' &   $\Lambda$   &   first kind  \\
		& P & first kind \\
		P-31m & $\Gamma-Z$ & first kind \\
		 & $K-H$ & first kind \\
		P-3m1 & $\Gamma-Z$ & first kind \\
		R-3m  & $\Lambda$ & first kind \\
		 & P & first kind \\
		P-61' & $\Gamma-Z$ & second kind \\
		P-6' & $\Gamma-Z$ & second kind \\
		P6'/m' & $\Gamma-Z$ & second kind \\
		P6mm & $K-H$ & first kind \\
		P6mm1' & $K-H$ & first kind \\
		P6'm'm & $\Gamma-Z$ & first kind \\
		P6'mm' &  $\Gamma-Z$    &  first kind \\
		& $K-H$ & first kind \\
		P-6m2 & $\Gamma-Z$ & first kind \\
		P-6m21' & $\Gamma-Z$ & first kind \\
		P-6'm'2 & $\Gamma-Z$ & second kind \\
		P-6'm2' & $\Gamma-Z$ & first kind \\
		P-62m & $\Gamma-Z$ & first kind \\
		 & $K-H$ & first kind \\
		P-62m1' & $\Gamma-Z$ & first kind \\
		& $K-H$ & first kind \\
		P-6'2'm & $\Gamma-Z$ & first kind \\
		& $K-H$ & first kind \\
		P-6'2m' & $\Gamma-Z$ & second kind \\
		P6/mmm & $K-H$ & first kind \\
		P6'/m'm'm &  $\Gamma-Z$ & first kind \\
		P6'/m'mm' & $\Gamma-Z$ & first kind \\
		& $K-H$ & first kind \\
		
		P-43m & 1 1 1 & first kind \\
		P-43m1' & 1 1 1 & first kind \\

		F-43m & 1 1 1 & first kind \\
		F-43m1' & 1 1 1 & first kind \\
		
		I-43m & 1 1 1 & first kind \\
		 & F & first kind \\
		I-43m1' & 1 1 1 & first kind \\
		 & F & first kind \\
		
		Pm3m & 1 1 1 & first kind \\
		Fm3m & 1 1 1 & first kind \\
		Im3m & 1 1 1 & first kind \\
		 & F & first kind \\
		\hline
		\hline
	\end{tabular}
	\label{tab:3}
\end{table}

\section{IV. The coexistence of Dirac point and triple point}
\begin{figure}
	\centering
	\includegraphics[width=5.5cm]{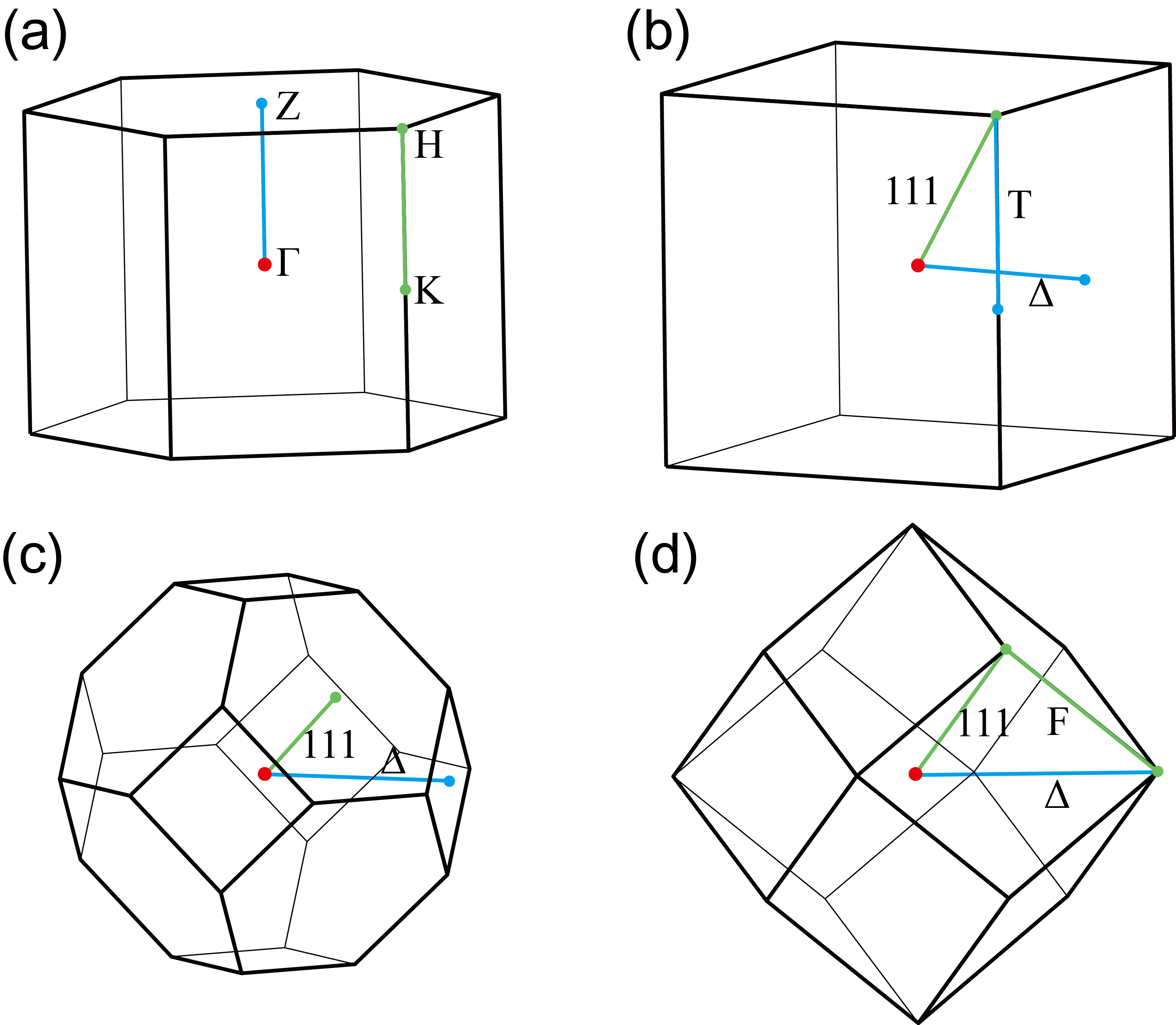}
	\caption{The Brillouin zone of the systems which allow the coexistence of Dirac points and triple points. The Brillouin zone of (a) P6/mmm, P6mm and P6mm1'. The Brillouin zone of (b) Pm3m, (c) Fm3m and (d) Im3m. The $k$ paths, which allow the existence of triple points, are marked in green. The $k$ paths, which allow the existence of Dirac points, are marked in blue.
	}\label{Fig: 2}
\end{figure}

The degeneracy of a Dirac fermion in high energy physics is protected by TRS and IS. However, the degeneracy of a Dirac fermion in condensed matter can be preserved in a $k$ path whose symmetry group has two or more than two 2-dimensional double group irreducible representations (as mentioned in section. II). That means Dirac fermion can exist in a system which does not contain (TRO $\cdot$ IS) symmetry. This gives rise to the possibility of the coexistence of Dirac fermion and odd-fold degenerate fermion. We know that triple points can exist in a $k$ path whose symmetry group contains both 1-dimensional and 2-dimensional double group irreducible representations (as mentioned in section. II too). Combining the condition of existence of Dirac fermion with the condition of existence of triple point, we find that there are several systems which allow the coexistence of Dirac points and triple points. $T$ and $\Delta$ of $Pm3m$, $\Delta$ of $Fm3m$, $\Delta$ of $Im3m$, $\Gamma-Z$ of $P6/mmm$, $\Gamma-Z$ of $P6mm$ and $\Gamma-Z$ of $P6mm{1}'$ allow the existence of Dirac points while the $k$ paths of these several systems, which allow the existence of triple points, are given in Table. \ref{tab:3}. The Brillouin zone of these several systems are shown in Fig. \ref{Fig: 2}.

Then there are two questions surfaced:

(1) Is the condition of existence of Weyl points similar to the condition of existence of triple points? More specifically, can Weyl points exist in a $k$ path whose symmetry group contains two 1-dimensional double group irreducible representations?

(2) Can the coexistence of Dirac points, triple points and Weyl points be allowed in some symmetric systems?

To answer question (1), we analyze two systems:
The first system has $D_{3d}$ point group symmetry with hexagonal Bravais lattice. The symmetry group of $\Gamma-Z$ in its Brillouin zone is $C_{3v}$. $C_{3v}$ contains two 1-dimensional double group irreducible representations. These two representations have different mirror symmetry eigenvalues ($i$ and $-i$) which means they are two different representations in the mirror plane of $C_{3v}$ symmetry. Thus the crossing point in $\Gamma-Z$ must extend onto the mirror planes of $C_{3v}$ symmetry to form a Weyl nodal line other than discrete points. This means, in this case, Weyl points are not able to exist in a $k$ path whose symmetry group contains two 1-dimensional double group irreducible representations.

The second system has $C_{3i}$ point group symmetry with hexagonal Bravais lattice. The symmetry group of $\Gamma-Z$ is $C_{3}$. $C_{3}$ contains two 1-dimensional double group irreducible representations. This system can be viewed as the first system to which a uniform magnetic field-$B_{z}$, in parallel with principle axis, is applied. The $B_{z}$ breaks the mirror symmetry of $C_{3v}$ and the Weyl nodal line is gap opening. Furthermore, the band structure along $k_{z}$ path can be described by Eq. \ref{eq:2} plus several $B_{z}$ induced terms:
\begin{equation} \label{eq:3}
	\begin{array}{lcl}
		H_{C_{3i}}(k_{z})=\varepsilon_{0}(k_{z})+ \\
				\begin{mpmatrix}
					M(k_{z})+\beta_{1} & 0 & 0 & 0 \\
					0 & M(k_{z})+\beta_{2} & 0 & 0 \\
					0 & 0 & -M(k_{z})+\beta_{3} & D+\beta_{5} \\
					0 & 0 & D\beta_{5}^{*} & -M(k_{z})+\beta_{4} \\
				\end{mpmatrix},
			\end{array}
\end{equation}
where $\beta_{1}, \ \beta_{2}, \ \beta_{3}$ and $\beta_{4}$ are real while $\beta_{5}$ is complex; $M(k_{z})$ is $C_{0}-C_{1}k^{2}_{z}$. Solving the eigenvalues of Eq. \ref{eq:3}, we find that the 2-dimensional double group irreducible representation-$\overline{E}_{1g/u}$ in Eq. \ref{eq:2} splits into two 1-dimensional double group irreducible representations. That means each triple point in Eq. \ref{eq:2} splits into two crossing points. Using $k \cdot P$ expansion and method of invariants to calculate the Hamiltonian around these crossing points can prove that these crossing points are Weyl points\cite{coex1}. Therefore, the two 1-dimensional double group irreducible representations of $\Gamma-Z$ in this system allow the existence of Weyl points.

According to the above analysis for the two systems, sometimes Weyl points can exist in a $k$ path whose symmetry group contains two 1-dimensional double group irreducible representations and sometimes Weyl points cannot. Thus, the condition of existence of Weyl points should not be as stated in question (1).

For answering question (2):

The $\Gamma-Z$ of the two above mentioned systems does not allow the coexistence of Dirac points, triple points and Weyl points, but recent research shows that triple point and Weyl point are able to coexist with each other\cite{in7-1}.

The symmetry requirement of Weyl points is much lower than that of triple points. Weyl points are robust against the perturbations of most of the symmetry breaking. Weyl points even can exist at the $k$ points whose symmetry group belongs to $C_{1}$\cite{coex2}. Such Weyl points in bulk of 3-dimensional systems only need periodic symmetry in the absence of (TRO $\cdot$ IS).

Therefore, in view of symmetry analysis hereinabove, the coexistence of Dirac points, triple points and Weyl points is allowed in some systems. If that happens, it will cause chaos when the systems are defined to be Dirac semimetal or triple point topological metal or Weyl semimetal. There is a viewpoint that triple point topological metal is an intermediate phase between Dirac and Weyl semimetal\cite{in7-4}. It can be true, but there are overlapping among these three phases and the boundary of the phases is still obscure so far. Such a state of chaos is raised by using symmetry to classify these three phases. In condensed matter systems, the symmetry group of a system is unique, but, in the Brillouin zone of a system, there are many different $k$ paths/planes belonging to different symmetry groups. Usually, the existences of Dirac points or triple points in a $k$ path depend on the symmetry group of the $k$ path other than only depend on the symmetry group of the system. As a result, the coexistence of Dirac points, triple points and Weyl points is symmetry-allowed.

\section{V. Conclusion}
In high energy physics, breaking TRS or breaking IS but keep Lorentz invariance, one Dirac fermion will split into two Weyl fermions. In condensed matter systems, breaking TRS or IS, Dirac point can remain intact or split into triple points or split into Weyl points. One Dirac point can split into two triple points, two triple points can split into four Weyl points. Existence of triple point between Dirac phase and Weyl phase was well hidden in the past. The recent discovery of triple point drives researchers into study of unknown characteristics of triple point. Therefore, we extend the theory of searching triple points to all symmorphic magnetic systems, and list all $k$ paths of all symmorphic systems which allow the existence of triple points. Our systematic study is helpful for searching triple points in various systems. Besides, we also find out that the coexistence of Dirac points and triple points is symmetrically allowed in some particular symmetric systems. Our works will not only be helpful for searching triple points but also extend the knowledge of such a new topological fermion.

	\begin{center}
		\textbf{\large ACKNOWLEDGMENTS}
	\end{center}
	
	This work is supported by the Ministry of Science and Technology of Taiwan under Grant No. MOST 104-2112-M-002-007-MY3.% We are grateful to Computer and Information Networking Center, National Taiwan University for the support of high-performance computing facilities.

	\section{APPENDIX: Corepresentations of black and white magnetic point group}
All black and white point groups can be express as follows:
		\begin{equation} \label{eq:A1}
M=H+TRO(G-H),
%			\begin{array}{lcl}
%				H_{D_{3d}}(k_{z})=\varepsilon(k_{z})+ \\
%				\begin{mpmatrix}
%					M_{0}-M_{1}k^{2}_{z} & 0 & 0 & 0 \\
%					0 & M_{0}-M_{1}k^{2}_{z} & 0 & 0 \\
%					0 & 0 & -M_{0}+M_{1}k^{2}_{z} & C \\
%					0 & 0 & C & -M_{0}+M_{1}k^{2}_{z} \\
%				\end{mpmatrix},
%			\end{array}
		\end{equation}
where $M$ is black and white point group, $H$ is the unitary subgroup of $M$ and $G$ is one of the ordinary point groups. 

Here, we denote element of $H$ by $U$, and element of $TRO(G-H)$ by $V$. We suppose that $\Delta$ is a unitary irreducible representation of $H$ (the dimension of $\Delta$ can be greater than 1), therefore:
		\begin{equation} \label{eq:A2}
		U\left\langle \varphi \right|=\left\langle \varphi \right|\Delta(U).
		\end{equation}
Now we introduce a basis $\left| \phi \right>$ which is produced by operating $V$ on $\left| \phi \right>$:
		\begin{equation} \label{eq:A3}
		V\left\langle \varphi \right|=\left\langle \phi \right|.
		\end{equation}
From Eq. \ref{eq:A2}, Eq. \ref{eq:A3} and $V^{-1}UV$ is belong to $H$, we can get:
		\begin{equation} \label{eq:A4}
\begin{array}{lcl}
		U\left\langle \phi \right|=UV\left\langle \varphi \right| \\
		 \ \ \ \ \ \ \ =V(V^{-1}UV)\left\langle \varphi \right| \\
		 \ \ \ \ \ \ \ =V\left\langle \varphi \right|\Delta(V^{-1}UV) \\
		 \ \ \ \ \ \ \ =\left\langle \phi \right|\Delta^{*}(V^{-1}UV),
\end{array}
		\end{equation}
complex conjugate is denoted by asterisk. Let
\begin{equation} \label{eq:A5}
\left\langle \zeta \right|=\left\langle \varphi, \phi \right|.
\end{equation}
From Eq. \ref{eq:A2}, Eq. \ref{eq:A4} and Eq. \ref{eq:A5}, we have:
\begin{equation} \label{eq:A6}
U\left\langle \zeta \right|=\left\langle \zeta \right|D(U),
\end{equation}
where
\begin{equation} \label{eq:A7}
D(U)=\begin{mpmatrix}
 \Delta(U) & 0 \\
  0 & \Delta^{*}(V^{-1}UV) \\
\end{mpmatrix},
\end{equation}
for all $U$ that belong to $H$. Since $\Delta^{*}(V^{-1}UV)$ is also a representation of $H$\cite{ap1}, the anti-unitary operators of $M$ do not create any extra irreducible representation. Anti-unitary operators only cause the irreducible representations of $H$ degenerate with each other or degenerate with itself, but usually anti-unitary operators do not cause any extra degeneracy.

If $\Delta(U)$ and $\Delta^{*}(V^{-1}UV)$ are equivalent, then there exists an unitary operator P such that:
\begin{equation} \label{eq:A8}
	\Delta(U)=P\Delta^{*}(V^{-1}UV)P^{-1},
\end{equation}
for all $U$ belonging to $H$.

If
\begin{equation} \label{eq:A9}
PP^{*}=\Delta(V^{2}),
\end{equation}
then anti-unitary operators do not cause any extra degeneracy, and we call it case(1).

If
\begin{equation} \label{eq:A10}
PP^{*}=-\Delta(V^{2}),
\end{equation}
then anti-unitary operators cause the irreducible representation $\Delta$ degenerate with itself, and we call it case(2).

If $\Delta(U)$ and $\Delta^{*}(V^{-1}UV)$ are not equivalent, then they are degenerate with each other, and we call it case(3).

Now we can classify all the representations of all black and white point groups. All those black and white point groups contain representations belonging to case(2) or case(3) are listed in Table. \ref{tab:S1}.

\begin{table}[h]
	\caption{The list for all the black and white magnetic point groups which have representations belonging to case(2) or case(3). The first column is the label of $M$. The second column is the label of $H$. The third column is the label of irreducible representation of $H$. The fourth column is the classification of the irreducible representation of $H$.}
	\begin{tabular}{c | c | c | c}
		\hline
		\hline
		$M$ & $H$ & irreducible representation of $H$ & case \\
		\hline
		-1' &  1 ($C_{1}$)  &   $\overline{A}$           &     2       \\
		2'/m &  m ($C_{1h}$)  &       ${}^{1}\overline{E}$, ${}^{2}\overline{E}$     &     3       \\
		2/m' &  2 ($C_{2}$)  &      ${}^{1}\overline{E}$, ${}^{2}\overline{E}$        &    3       \\
		4' &  2 ($C_{2}$)  &     ${}^{1}\overline{E}$, ${}^{2}\overline{E}$         &    3       \\
		-4' & 2 ($C_{2}$)   &     ${}^{1}\overline{E}$, ${}^{2}\overline{E}$          &     3       \\
		4'/m &  2/m ($C_{2h}$)  &     ${}^{1}\overline{E}_{g}$, ${}^{2}\overline{E}_{g}$, ${}^{1}\overline{E}_{u}$, ${}^{2}\overline{E}_{u}$        &     3       \\
		4/m' &  4 ($C_{4}$)  &        ${}^{1}\overline{E}_{2}$, ${}^{2}\overline{E}_{1}$, ${}^{1}\overline{E}_{1}$, ${}^{2}\overline{E}_{2}$      &    3       \\
		4'/m' &  -4 ($S_{4}$)  &     ${}^{1}\overline{E}_{2}$, ${}^{2}\overline{E}_{1}$, ${}^{1}\overline{E}_{1}$, ${}^{2}\overline{E}_{2}$        &    3       \\
		-3' &  3 ($C_{3}$)  &   $\overline{A}$           &     2       \\
            &               &   ${}^{1}\overline{E}$, ${}^{2}\overline{E}$  &   3  \\
		-3'm &  3m ($C_{3v}$)  &   ${}^{1}\overline{E}$, ${}^{2}\overline{E}$            &     3       \\
             &                 &   $\overline{E}_{1}$  &  1   \\
		-3'm' &  32 ($D_{3}$)  &   ${}^{1}\overline{E}$, ${}^{2}\overline{E}$            &     3       \\
		      &                 &   $\overline{E}_{1}$  &  1   \\
		6' &  3 ($C_{3}$)  &   ${}^{1}\overline{E}$, ${}^{2}\overline{E}$           &     3       \\
		   &                 &   $\overline{A}$  &  1   \\
		-6' &  3 ($C_{3}$)  &  ${}^{1}\overline{E}$, ${}^{2}\overline{E}$           &     3       \\
		    &                 &   $\overline{A}$  &  1   \\
		6'/m &  -6 ($C_{3h}$)  &     ${}^{1}\overline{E}_{1}$, ${}^{2}\overline{E}_{3}$, ${}^{1}\overline{E}_{2}$, ${}^{2}\overline{E}_{2}$, ${}^{2}\overline{E}_{1}$, ${}^{1}\overline{E}_{3}$        &     3       \\
		6/m' &  6 ($C_{6}$)  &     ${}^{1}\overline{E}_{1}$, ${}^{2}\overline{E}_{3}$, ${}^{1}\overline{E}_{2}$, ${}^{2}\overline{E}_{2}$, ${}^{2}\overline{E}_{1}$, ${}^{1}\overline{E}_{3}$         &     3       \\
		6'/m' &  -3 ($C_{3i}$)  &   ${}^{1}\overline{E}_{g}$, ${}^{2}\overline{E}_{g}$, ${}^{1}\overline{E}_{u}$, ${}^{2}\overline{E}_{u}$           &     3       \\
		    &                 &   $\overline{A}_{g}$, $\overline{A}_{u}$  &  1   \\
		m'3 &  23 ($T$)  &     ${}^{1}\overline{F}$, ${}^{2}\overline{F}$       &     3       \\
		    &                 &   $\overline{E}$  &  1   \\
		\hline
		\hline
	\end{tabular}
	\label{tab:S1}
\end{table}

\end{document}